\documentclass[twocolumn,preprintnumbers,amsmath,amssymb,pra,a4paper]{revtex4-1}

\usepackage{amssymb}
\usepackage{amsfonts}
\usepackage{amsmath}
\usepackage{amsthm}
\usepackage[pdftex]{graphicx}
\usepackage{braket}
\usepackage{bm}

\newcommand{\trans}[1]{{#1}^{\ensuremath{\mathsf{T}}}} 

\begin{document}

\title{Gaussian matrix-product states for coding in bosonic communication channels}
\author{Joachim Sch\"afer} 
\author{Evgueni Karpov} 
\author{Nicolas J. Cerf} 
\affiliation{QuIC, Ecole Polytechnique de Bruxelles, CP 165/59, Universit\'e Libre de Bruxelles, B-1050 Brussels, Belgium}
	
\begin{abstract}
The communication capacity of Gaussian bosonic channels with memory has recently attracted much interest. Here, we investigate
a method to prepare the multimode entangled input symbol states for encoding classical information into these 
channels. In particular, we study the usefulness of a Gaussian matrix-product state (GMPS) as an input symbol state, 
which can be sequentially generated although it remains heavily entangled for an arbitrary number of modes. 
We show that the GMPS can achieve more than 99.9\% of the Gaussian capacity for Gaussian bosonic memory channels
with a Markovian or non-Markovian correlated noise model in a large range of noise correlation strengths. 
Furthermore, we present a noise class for which the GMPS is the \emph{exact} optimal input symbol state of the corresponding channel. 
Since GMPS are ground states of particular quadratic Hamiltonians, our results suggest a possible link between the theory of quantum communication channels and quantum many-body physics.
\end{abstract}

\maketitle
\section{Introduction}

Quantum communication channels are at the heart of quantum information theory. Among quantum channels, the bosonic Gaussian channels describe very common physical links, such as the transmission via free space or optical fibers \cite{CD94}. The fundamental feature of a quantum channel is its capacity, which is the maximal information transmission rate for a given available energy. The capacity can be  classical or quantum, depending on whether one sends classical or quantum information (here, we focus on the former). In previous works, it was shown that for certain quantum \emph{memory} channels, in particular channels with correlated noise, the optimal input symbol state is entangled across successive uses of the channel; see Refs. \cite{MemChan,CCMR05,GM05,BDM05,KDC06,PZM08,LPM09,PLM09,SDKC09,SKC11} and references therein. In general, such multimode entangled states may be quite hard to prepare, which motivates the present work.

In this paper, we address the problem of implementing the (optimal) input symbol state for Gaussian bosonic channels with particular memory models. For this purpose, we study the usefulness of the so-called Gaussian matrix-product state (GMPS) \cite{SCW06,AE06} as an input symbol state for the Gaussian bosonic channel with additive noise \cite{SDKC09,SKC11} and the lossy Gaussian bosonic channel \cite{PZM08,LPM09,PLM09}. This translationary-invariant state is heavily entangled and can be generated sequentially, which happens to be crucial for its use as a multimode input symbol state in the transmission via a Gaussian bosonic channel. The GMPS are known to be a useful resource for quantum teleportation protocols \cite{LB00,AI05}, but, to our knowledge, they have never been considered in the context  of quantum channels.

In Sec. II, we give an overview of the method used to derive the Gaussian capacity of Gaussian bosonic memory channels, following our previous work \cite{CCMR05,SDKC09,SKC11}. Our original results are presented in Sec. III, where we address the use of GMPS in this context.
In Sec. III A, we show that the GMPS, though not being the optimal input state, is close-to-capacity achieving for Gaussian bosonic channels with a Markovian and non-Markovian noise in a large region of noise correlation strengths. In Sec. III B, we provide a class of noisy channels for which the GMPS is the {\it exact} optimal input state. Since the GMPS is as well the ground state of particular quadratic Hamiltonians, this suggests a direct link between the maximization of information transmission in quantum channels and the energy minimization of quantum many-body systems.
In Sec. III C, we also observe that the squeezing strengths that are needed to realize the GMPS in an optical setup are experimentally feasible. Finally, our conclusions are provided in Sec. IV. 

\section{Gaussian capacity of memory channels with correlated noise}
\subsection{Gaussian bosonic channels}
Let us now consider an $n$-mode optical channel $T^{(n)}$, which can either be a bosonic additive noise channel or a lossy bosonic channel. In the following, $n$ single-mode channel uses will be equivalent to one use of an $n$-mode parallel channel \cite{SW97}. Each mode $j$ is associated with the annihilation $\hat{a}_j$ and creation $\hat{a}_j^{\dagger}$ operators, or equivalently with the pair of quadrature operators 
$\hat{q}_j = (\hat{a}_j + \hat{a}_j^{\dagger})/\sqrt{2}$ and $\hat{p}_j = i(\hat{a}_j^{\dagger} - \hat{a}_j)/\sqrt{2}$, which obey the canonical commutation relation $[\hat{q}_i,\hat{p}_j] = i\delta_{ij}$.
By defining the vector of quadratures $\bm{\hat{R}} = (\hat{q}_1,...,\hat{q}_n;\hat{p}_1,...,\hat{p}_n)^\mathrm{T}$, we can express the displacement vector $\bm{m} = \mathrm{Tr}{[ \, \rho  \bm{\hat{R}}]}$ of any state $\rho$, along with its covariance matrix (CM)
\begin{equation}\nonumber
	\begin{split}
		\bm{\gamma} & = \mathrm{Tr}{[ (\bm{\hat{R}} - \bm{m}) \, \rho \, (\bm{\hat{R}} - \bm{m})^{\ensuremath{\mathsf{T}}} ]} - \bm{J}/2,\\
		\textrm{with} \quad \bm{J} & = i \begin{pmatrix}0 & \bm{I}\\-\bm{I} & 0\end{pmatrix},
	\end{split}
\end{equation}
where $\bm{I}$ is the $n \times n$ identity matrix. In phase space, a Gaussian state is defined as a state $\rho$ having a Wigner distribution that is Gaussian; hence, it is fully characterized by its mean $\bm{m}$ and CM $\bm{\gamma}$. 

For the channel encoding, we consider a continuous alphabet, that is, we encode a complex number $q+ip$ instead of a discrete index into each symbol state. We encode a message of length $n$ into a $2n$-dimensional real vector $\bm{r} =  (q_1,q_2,...,q_n;p_1,p_2,...,p_n)^{\ensuremath{\mathsf{T}}}$. Physically, this encoding corresponds in phase space to a displacement by $\bm{r} $  
of the $n$-partite Gaussian input state defined by its mean $\bm{m}_{\rm in}$ and CM $\bm{\gamma}_{\rm in}$. The modulation of the multipartite input state is taken as a (classical) Gaussian multipartite probability density $p_\mathrm{mod}(\bm{r})$ with mean $\bm{m}_\mathrm{mod}$ and CM $\bm{\gamma}_\mathrm{mod}$. The means of the input state $\bm{m}_{\rm in}$ and classical modulation $\bm{m}_\mathrm{mod}$ can be set to zero without loss of generality because displacements leave the entropy invariant; hence, they do not play any role in the capacity formulas defined in Sec. II B. The action of the channel $T^{(n)}$ is thus fully characterized in terms of covariance matrices, that is,
\begin{equation}\label{eq:addchannel}
  \begin{split}
    & \bm{\gamma}_\mathrm{out} = \kappa \, \bm{\gamma}_\mathrm{in} + \kappa' \, \bm{\gamma}_\mathrm{env},\\
    & \overline{\bm{\gamma}}   = \bm{\gamma}_\mathrm{out} + \kappa \, \bm{\gamma}_\mathrm{mod},
  \end{split}
\end{equation}
where $\bm{\gamma}_\mathrm{out}$ and $\overline{\bm{\gamma}}$ are the CM of the individual output and modulated output states, respectively. For $\kappa=\kappa'=1$, Eq.~\eqref{eq:addchannel} defines the bosonic Gaussian channel with additive noise, where $\bm{\gamma}_{\rm env}$ is the CM of a (classical) Gaussian multipartite probability density $p_\mathrm{env}(\bm{r} )$ describing noise-induced displacements in phase space (see Ref. \cite{SKC11} for details). For $\kappa=\eta$ and $\kappa'=1-\eta$, with a beamsplitter transmittance $\eta \in [0,1]$, Eq.~\eqref{eq:addchannel} defines the lossy channel where $\bm{\gamma}_{\rm env}$ stands for the CM of the environment state (see Ref. \cite{PLM09} for details).
Both channels obey the physical energy constraint that reads $\mathrm{Tr}(\bm{\gamma}_{\rm in} + \bm{\gamma}_{\rm mod})/(2n) - 1/2 = \overline{n}$, where $\overline{n}$ is the mean photon number at the input.

\subsection{Gaussian capacity}
In recent works, we found the Gaussian capacity (i.e., the capacity when restricted to Gaussian input states according to the usual Gaussian channel minimum entropy conjecture) and optimal input encoding for the additive Gaussian channel with noise correlations between subsequent uses of the channel modeled by the CM \cite{SDKC09,SKC11}
\begin{equation}\label{eq:envcov}
	\bm{\gamma}_{\rm env} = 
	\begin{pmatrix}
		\bm{\gamma}_{\rm env}^q & 0\\
		0 & \bm{\gamma}_{\rm env}^p
	\end{pmatrix}
\end{equation}
where $\bm{\gamma}_{\rm env}^q$ and $\bm{\gamma}_{\rm env}^p$ are commuting matrices of dimension $n \times n$. The absence of correlations between $q$ and $p$ in Eq.~\eqref{eq:envcov} is generally considered to describe a natural noise. We found that the optimal input and modulation CM $\bm{\gamma}_{\rm in}^*$ and $\bm{\gamma}_{\rm mod}^*$ are diagonal in the same basis as the noise CM $\bm{\gamma}_{\rm env}$, and have the same block structure. Thus, $\bm{\gamma}_{\rm in}^* = {\bm{\gamma}_{\rm in}^{q*}} \oplus {\bm{\gamma}_{\rm in}^{p*}}$ and $\bm{\gamma}_{\rm mod}^* = {\bm{\gamma}_{\rm mod}^{q*}} \oplus {\bm{\gamma}_{\rm mod}^{p*}}$. In addition, the optimal input state is pure, i.e.,
$\det{(2\bm{\gamma}^*_{\rm in})}=1$, which implies
\begin{equation}\label{eq:optincm}
	\bm{\gamma}_{\rm in}^* = 
	\begin{pmatrix}
		{\bm{\gamma}_{\rm in}^{q*}} & 0\\
		0 & \frac{1}{4}({\bm{\gamma}_{\rm in}^{q*}})^{-1}
	\end{pmatrix}.
\end{equation}
From now on, we consider the optimal input and modulation eigenvalue spectra in the limit of an infinite number of channel uses $n \to \infty$, so all matrices must be expanded to infinite dimensions, see Ref. \cite{SDKC09}.

For an input energy $\overline{n}$ above a certain threshold $\overline{n}_{\rm thr}$, the optimal eigenvalue spectra are linked via a global \emph{quantum water filling} solution \cite{SDKC09}, that is, $\overline{\gamma}^{q*}(x)~=~\overline{\gamma}^{p*}(x)~=~\textrm{const.}, \forall x \in {\mathcal A}$ where $x$ is a continuous spectral parameter within a spectral domain ${\mathcal A}$ and $\overline{\gamma}^{q,p*}(x)$ is the spectrum of the $q$ and $p$ blocks of the optimal modulated output CM $\overline{\bm{\gamma}}^* = \overline{\bm{\gamma}}^{q*} \oplus \overline{\bm{\gamma}}^{p*}$. Furthermore, the optimal input state was determined as \cite{SDKC09,PLM09} 
\begin{equation}\label{eq:optin}
	{\gamma_{\rm in}^{q,p*}}(x) = \frac{1}{2}\sqrt{\frac{\gamma_{\rm env}^{q,p}(x)}{\gamma_{\rm env}^{p,q}(x)}},
\end{equation}
which corresponds more precisely to the spectrum of the $q$ and $p$ blocks of the optimal input CM, $\bm{\gamma}_{\rm in}^*$. 
We remark that this holds for both the additive noise \cite{SDKC09} and lossy channel \cite{PLM09}. 

In the following, we will consider noise models (see Sec. II C) characterized by a CM with symmetric spectrum, i.e., $\gamma_{\rm env}^q(x)=\gamma_{\rm env}^p(|{\mathcal A}|-x)$, where $|{\mathcal A}|$ is the size of the spectral domain ${\mathcal A}$. Furthermore, the noise models fulfill $\max_{x}\{ \gamma_{\rm env}^q(x) \} = \gamma_{\rm env}^q(0)$. For this case, the input energy $\overline{n}$ that is required to fulfill the global quantum water filling solution and Eq.~\eqref{eq:optin}, for all $x$, is given by
\begin{equation}\label{eq:nthr}
	\overline{n} \geq \overline{n}_{\rm thr} \equiv  {\gamma_{\rm in}^{q*}}(0) + \gamma_{\rm env}^{q}(0)  - \frac{1}{2} - \bar{N},
\end{equation}
where $\bar{N}=\frac{1}{|{\mathcal A}|}\int_{x \in {\mathcal A}}dx \, \gamma_{\rm env}^q(x)$ stands for the added noise energy. Throughout this paper, we only consider the case above threshold, when $\overline{n}~\geq~\overline{n}_{\rm thr}$. Then, the Gaussian capacity of the channel with additive noise is given by \cite{SKC11}
\begin{eqnarray}
	C = & & g\left(\overline{n} + \bar{N}\right)\nonumber\\
	    & - & \frac{1}{|{\mathcal A}|}\int\limits_{x \in {\mathcal A}}{dx  \, g\left(\sqrt{\gamma_{\rm out}^{q*}(x)\gamma_{\rm out}^{p*}(x)} - \frac{1}{2}\right)}\label{eq:capglwf}, 
\end{eqnarray}
where $\gamma_{\rm out}^{q,p*}(x) = {\gamma_{\rm in}^{q,p}}^*(x)+\gamma_{\rm env}^{q,p}(x)$ according to Eq.~(\ref{eq:addchannel}). The function $g(x)$ stands for the entropy of a thermal state with $x$ photons. It is defined as $g(x) = (x+1)\log{(x+1)} - x\log{x}$ if $ x > 0$, and $g(x) = 0$ if $x \leq 0$, where $\log(x)$ denotes the logarithm to base 2.

Now, if one restricts the input states to independent coherent states in the case of global water filling, one may also define the 
\emph{coherent-state rate} \cite{SKC11}, which is given by Eq. \eqref{eq:capglwf} replacing ${\gamma_{\rm out}^{q,p*}}(x)$ by $1/2+\gamma_{\rm env}^{q,p}(x)$. For the lossy channel the quantities given by Eqs.~\eqref{eq:nthr} and \eqref{eq:capglwf} as well as the coherent-state rate are obtained by replacing $\overline{n} \rightarrow \eta \overline{n}, \gamma_{\rm env}^{q,p}(x) \rightarrow (1-\eta) \gamma_{\rm env}^{q,p}(x)$ and ${\gamma_{\rm in}^{q,p}}^*(x) \rightarrow \eta {\gamma_{\rm in}^{q,p}}^*(x)$. Note that all these expressions also rely on the assumption that  the Gaussian capacity of independent Gaussian channels is additive, see Ref. \cite{H05}.

\subsection{Noise models}\label{sec:noises}
Let us introduce two different noise models which will be used to model the Gaussian memory channels, 
namely a Markovian and non-Markovian model.

\subsubsection{Markov additive noise}
In Refs. \cite{SDKC09,SKC11}, we considered a classical Markov noise with variance $N_M \geq 0$, given by 
\begin{equation}\label{eq:envmark}
	{\bm{\gamma}_{\rm env}}_M = N_M
	\begin{pmatrix}
		\bm{M}(\phi) & 0\\
		0 & \bm{M}(-\phi)
	\end{pmatrix},
\end{equation}
where $\bm{M}(\phi)$ is an $n \times n$ matrix defined as $M_{ij}(\phi) = \phi^{|i-j|}$, with the correlation parameter $0 \leq \phi < 1$. Note that $\bm{M}(\phi)$ and $\bm{M}(-\phi)$ commute in the limit of an infinite number of channel uses.
In this limit, the spectra of the quadrature blocks ${\bm{\gamma}_{\rm env}^{q,p}}_M \equiv N_M\bm{M}(\pm \phi)$ are given by
\begin{equation}\label{eq:specM}
	{\gamma_{\rm env}^{q,p}}_M(x) = N_M \, \frac{1 - \phi^2}{1 + \phi^2 \mp 2\phi \cos(x)}, \quad x \in [0,2\pi],
\end{equation}
with the upper (lower) sign standing for the $q$ ($p$) quadrature. By using Eq.~\eqref{eq:optin}, we find that the optimal input state is an infinite  product of squeezed states. Then, when rotated back to its original basis, the optimal input state becomes a multimode entangled state \cite{SDKC09,SKC11}.
\subsubsection{Non-Markovian noise}
A non-Markovian channel noise model was considered in Refs. \cite{PLM09,LPM09}, given by
\begin{equation}\label{eq:envoleg}
	{\bm{\gamma}_{\rm env}}_{N} = N_{N}
	\begin{pmatrix}
		e^{s\bm{\Omega}} & 0\\
		0 & e^{-s\bm{\Omega}}
	\end{pmatrix},
\end{equation}
with $N_{N} \geq 1/2$ for the considered lossy channel ($N_{N} \geq 0$ for a non-Markovian additive noise channel), $s \in \mathbb{R}$, and where $\bm{\Omega}$ is a $n \times n$ matrix defined as $\Omega_{ij} = \delta_{i,j+1} + \delta_{i+1,j}$. The spectra of the quadrature blocks ${\bm{\gamma}_{\rm env}^{q,p}}_N \equiv N_N e^{\pm s \bm{\Omega}}$ read
\begin{equation}
	{\gamma_{\rm env}^{q,p}}_{N}(x) = N_{N} \, e^{\pm 2s \cos(x)}, \quad x \in [0,2\pi],
\end{equation}
with the upper (lower) sign standing for the $q$ ($p$) quadrature. 
In the case of a global water filling, it was shown that the optimal input state [Eq. \eqref{eq:optin}] is also entangled in the original basis \cite{LPM09,PLM09}, as for the Markov additive noise. 

Since the optimal input state for both noise models exhibits multimode entanglement across the subsequent uses of the channel, with $n\to\infty$, its preparation may be a very challenging task. This is what we investigate in the next section.

\section{Gaussian Matrix Product State}\label{sec:gmps}
\begin{figure}
	\centering
		\includegraphics[width=0.45\textwidth]{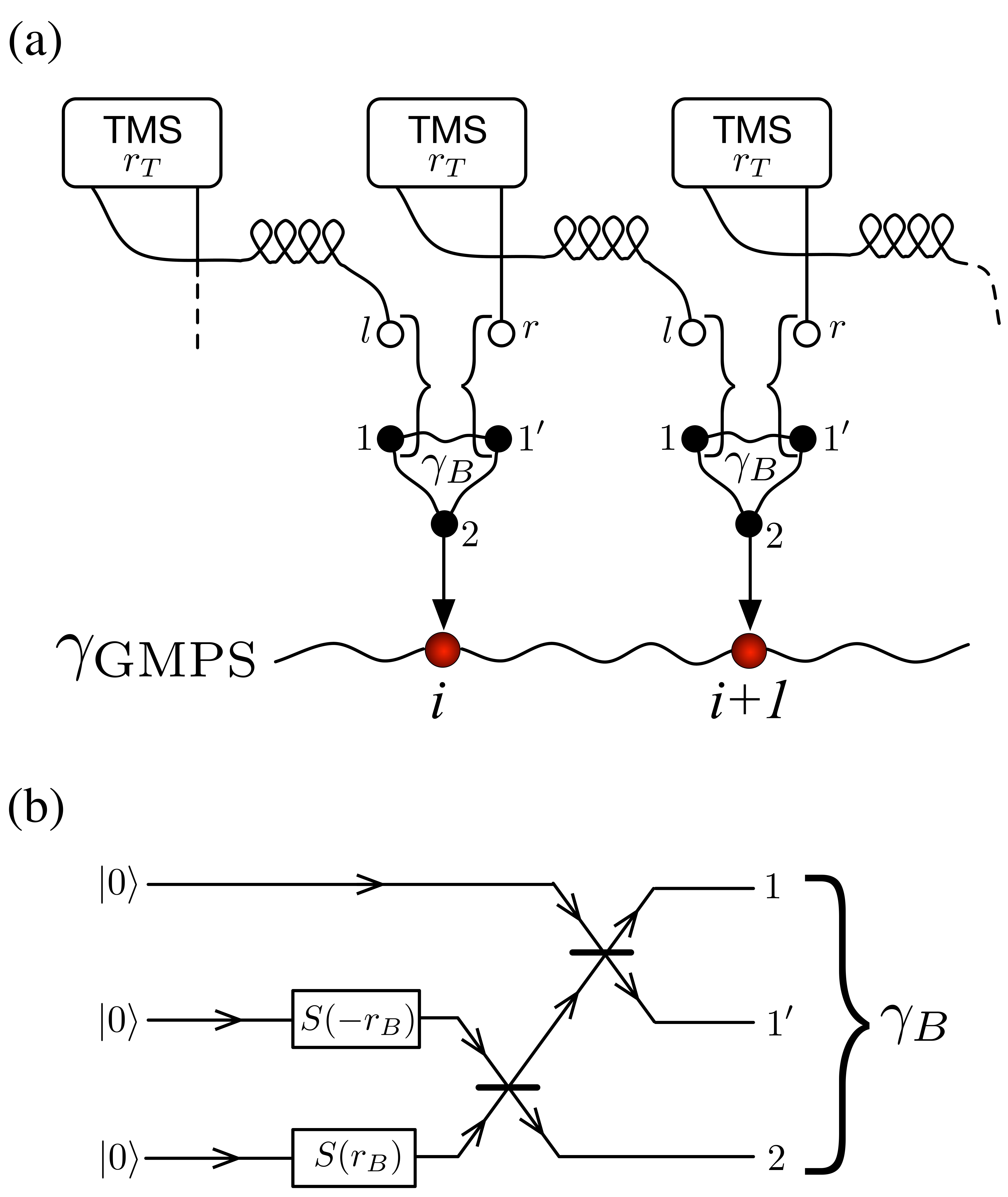}
	\caption{(a) Optical scheme for the preparation of the Gaussian matrix-product state (GMPS), slightly modified with respect to Ref. \cite{AE06}. 
Here, TMS stands for a two-mode squeezed vacuum state with squeezing $r_T$, while $\bm{\gamma}_B$ represents the three-mode building block. Note that all TMS and three-mode building blocks could each be generated by a single device that is used repeatedly. One half of the TMS generated at time $i$ is used immediately to generate the GMPS mode $i$, while the other half is sent to a delay line (to be used at time $i+1$). After two Bell measurements (represented by curly brackets) involving the two TMS halves (noted $l$ and $r$) and the two upper modes of $\bm{\gamma}_B$ (noted 1 and 1') followed by appropriate conditional displacements, the third mode (noted 2) of $\bm{\gamma}_B$ collapses into the GMPS mode $i$. (b) Optical setup of the three-mode building block $\bm{\gamma}_B$ that is used to generate a nearest-neighbor correlated GMPS. Here $\ket{0}$ denote vacuum modes, $S(r_{B})$ is a one-mode squeezer with parameter $r_{B}$, and the bold horizontal bars represent $50{\rm:}50$ beamsplitters.}
	\label{fig:gmps}
\end{figure}

We now address the question of how to optically implement the optimal input states. In this context, we examine the so-called Gaussian matrix-product state (GMPS), which is heavily entangled just as the optimal input state, has a known optical implementation, and can be generated sequentially. This state was first discussed in Ref. \cite{SCW06} as the ground state of particular Hamiltonians of harmonic lattices. In general, GMPS are constructed by taking a fixed number $\mathcal{M}$ of finitely or infinitely entangled two-mode squeezed vacuum states shared by adjacent sites, and applying an arbitrary $2\mathcal{M}$ to 1 mode Gaussian operation on each site $i$. 

In what follows, we restrict our discussion to a pure, translationally-invariant, one-dimensional GMPS, and, furthermore, to a single finitely entangled two-mode squeezed (TMS) vacuum state per bond between adjacent sites $(\mathcal{M}=1)$. We use the protocol introduced in Ref. \cite{AE06}, depicted in a slightly modified form in Fig.~\ref{fig:gmps}(a). Each GMPS mode $i$ is obtained by operating on a three-mode entangled state (called ``building block''; see Refs. \cite{AE06,AE07} for details) together with the shares ($l$ and $r$) of the two TMS vacuum states 
connecting site $i$ to the left and right sites, respectively.
As shown in Fig.~\ref{fig:gmps}(a), a first teleportation is performed by making a Bell measurement on modes $l$ and $1$, followed by a conditional displacement on mode $1'$. A second teleportation then is made with a Bell measurement on modes $r$ and $1'$, followed by a   
conditional displacement on mode $2$. The final state of mode $2$ then reduces precisely to that of the $i$th mode of the desired GMPS.
We focus now on the mathematical description of the GMPS and its use as an input state, while we discuss its experimental realization with single-mode squeezers in Sec. \ref{sec:experiment}.
\subsection{GMPS as approximating input state}
The CM of the GMPS can be written as 
\begin{equation}\label{eq:gmps}
	\bm{\gamma}_{\rm GMPS} = \frac{1}{2}
	\begin{pmatrix}
		\bm{\mathcal{C}}^{-1} & 0\\
		0 & \bm{\mathcal{C}}
	\end{pmatrix},
\end{equation}
where $\bm{\mathcal{C}}$ is a $n \times n$ circulant symmetric matrix. In Ref. \cite{SCW06}, it was proven that the correlations of a one-dimensional GMPS decay exponentially. Therefore, in the limit $n~\to~\infty$, the spectrum of $\bm{\mathcal{C}}^{-1}$ reduces (up to a change of variance) to the spectrum of $\bm{M}(\phi)$ \footnote{In the limit $n \rightarrow \infty$, the spectrum of a symmetric circulant matrix tends to the spectrum of its corresponding symmetric Toeplitz matrix (R. M. Gray, Found. Trends Commun. Inf. Theory {\bf 2}, 155 (2006).)}, that is,
\begin{eqnarray}
	\frac{1}{2}\lambda^{(\bm{\mathcal{C}}^{-1})}(x) & \equiv & \gamma_{\rm GMPS}^q(x)\nonumber\\
	& = & \tilde{N} \, \left( \frac{1 - \phi_{\rm in}^2}{1 + \phi_{\rm in}^2 - 2\phi_{\rm in} \cos(x)} + \Delta \right),\label{eq:gmpsq}
\end{eqnarray}
with $x \in [0,2\pi]$, $\tilde{N} \geq 0$, $0 \leq \phi_{\rm in} < 1$, $\Delta \in \mathbb{R}$, and the additional condition $\Delta \tilde{N} \geq -1/2$ ensuring that the spectrum corresponds to a quantum state.
\begin{figure}
	\centering
	\includegraphics[width=0.5\textwidth]{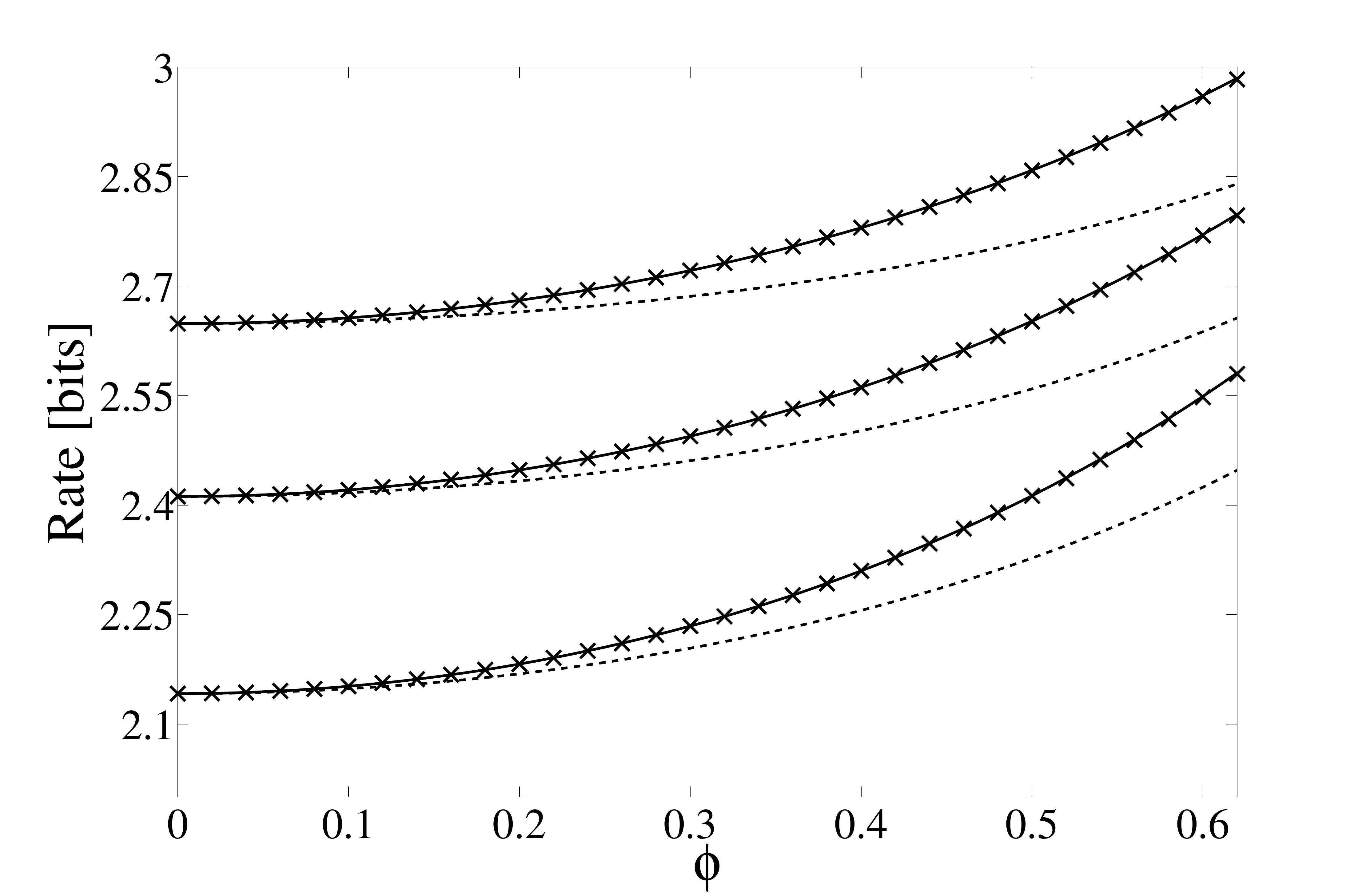}
	\caption{Rates of a channel with additive Markov noise: Gaussian capacity $C$ (solid line), GMPS-rate $R_{\rm GMPS}$ (crosses) and coherent-state rate $R_{\rm coh}$ (dashed line) vs. correlation $\phi$, where from top to bottom $N_M = \{0.5,0.7,1\}$. We took $\overline{n}=5$.}
	\label{fig:markrate}
\end{figure}
\begin{figure}
	\centering
	\includegraphics[width=0.5\textwidth]{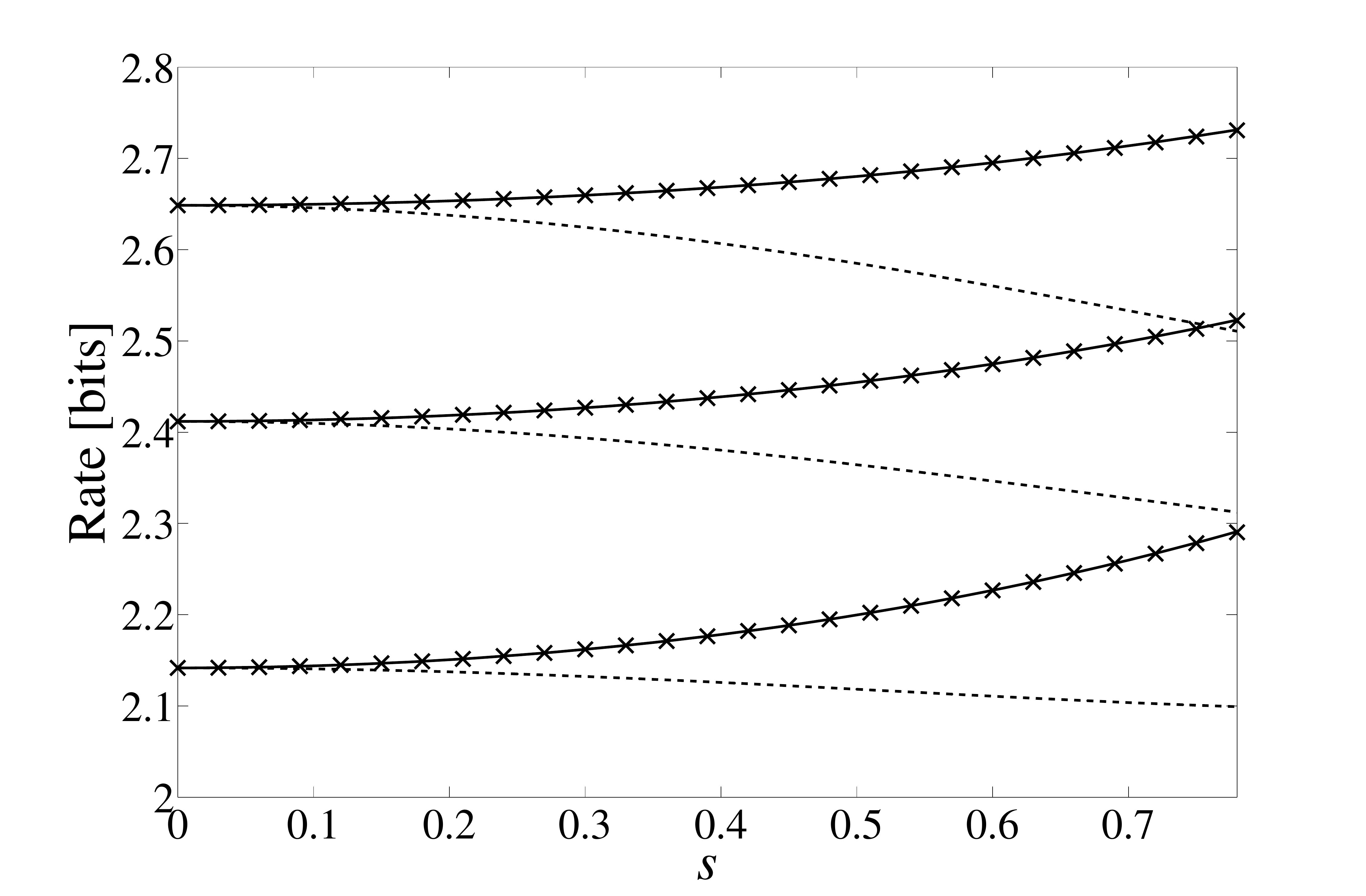}
	\caption{Rates of a channel with additive non-Markovian noise: Gaussian capacity $C$ (solid line), GMPS-rate $R_{\rm GMPS}$ (crosses) and coherent-state rate $R_{\rm coh}$ (dashed line) vs. correlation $s$, where from top to bottom $N_N = \{0.5,0.7,1\}$. We took $\overline{n}=5$.}
	\label{fig:nonmarkrate}
\end{figure}
\begin{figure}
	\centering
	\includegraphics[width=0.5\textwidth]{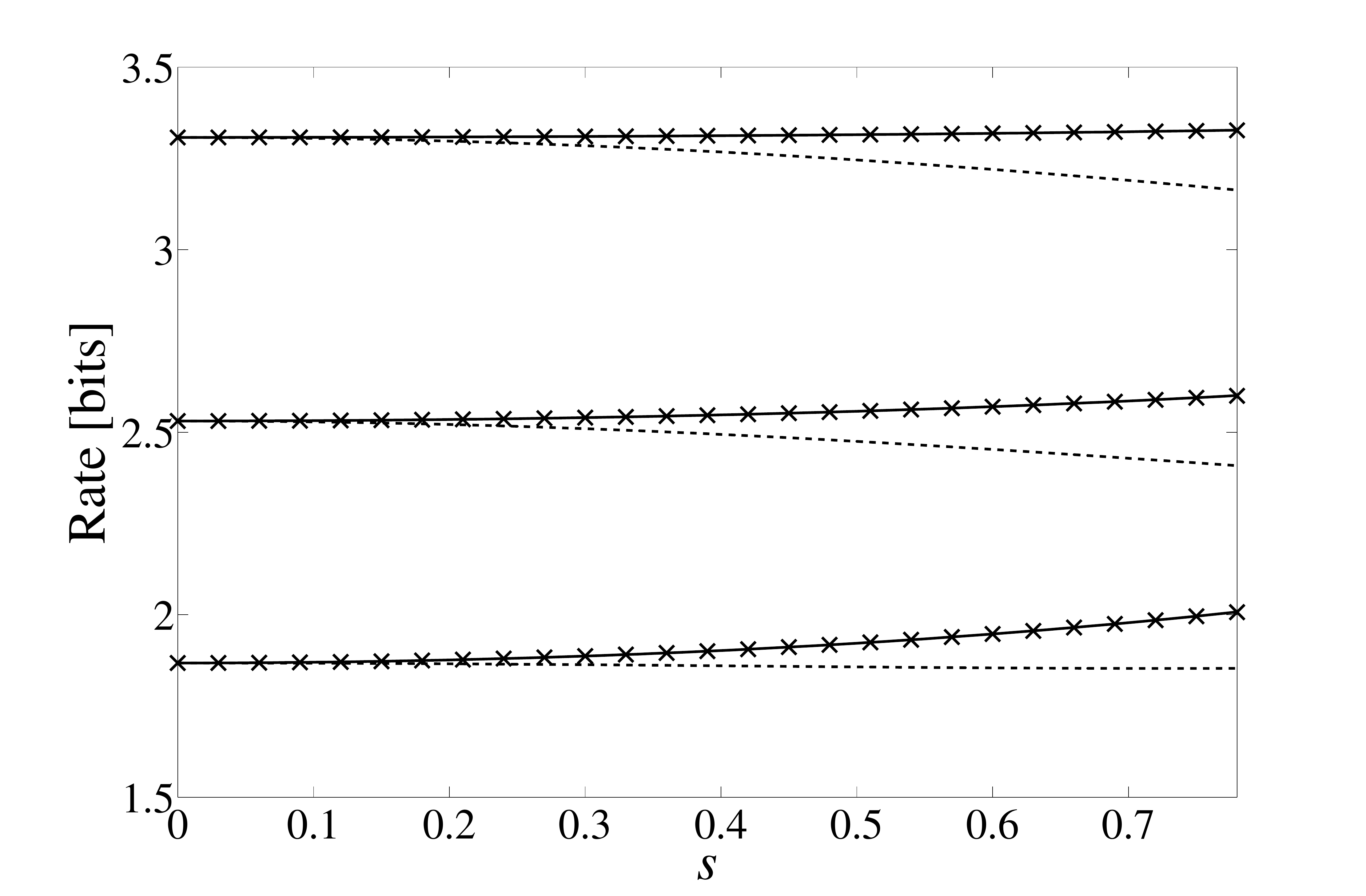}
	\caption{Rates of a lossy channel with non-Markovian noise: Gaussian capacity $C$ (solid line), GMPS-rate $R_{\rm GMPS}$ (crosses) and coherent-state rate $R_{\rm coh}$ (dashed line) vs. correlation $s$, where from top to bottom $\eta = \{0.5,0.7,0.9\}$. We took $N_N=1$ and $\overline{n}=5$.}
	\label{fig:nonmarklossyrate}
\end{figure}
\begin{figure}
	\begin{minipage}[t]{4.26cm} 
		\includegraphics[width=1\textwidth]{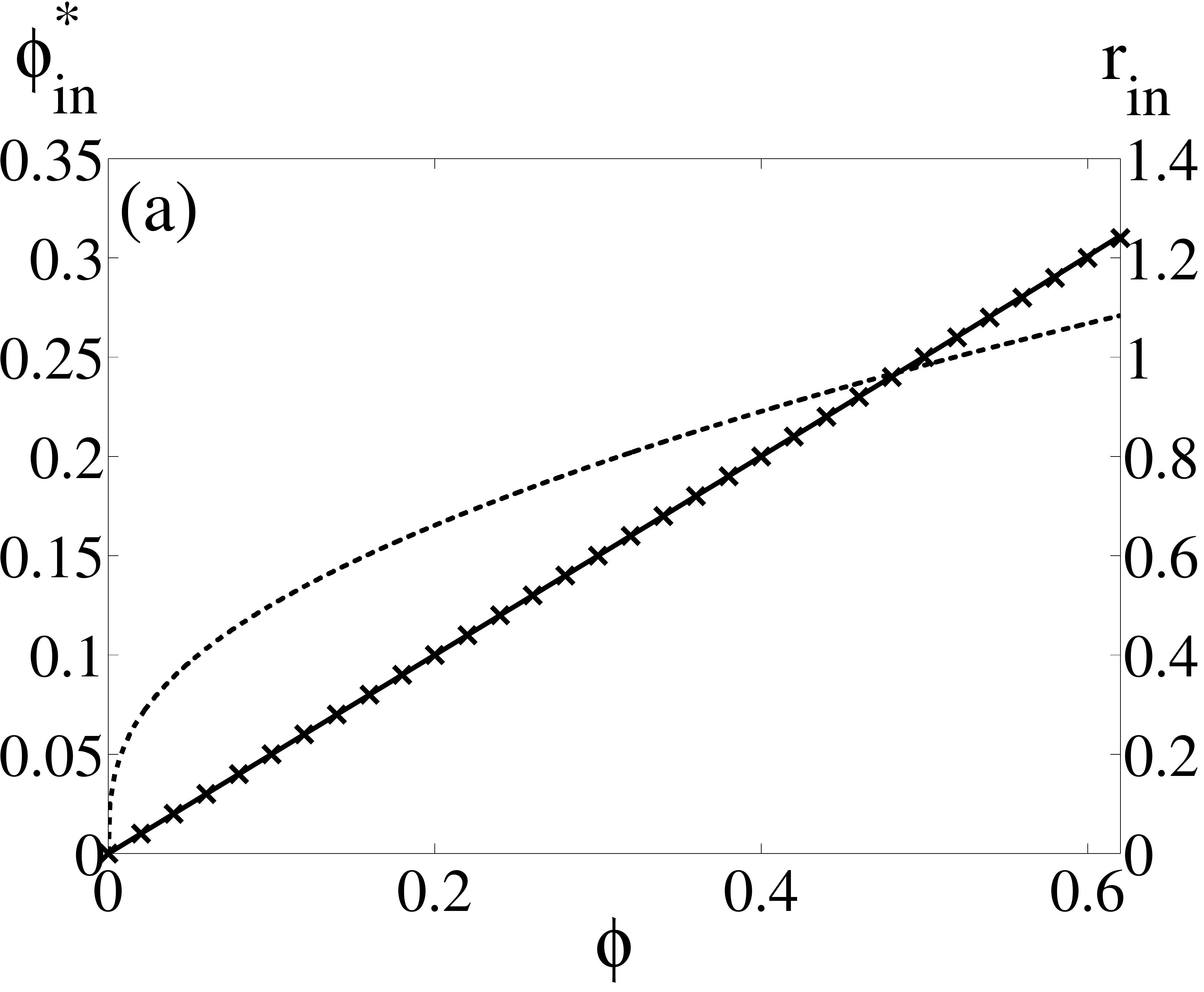} 
	\end{minipage} 
	\begin{minipage}[t]{4.26cm} 
		\includegraphics[width=1\textwidth]{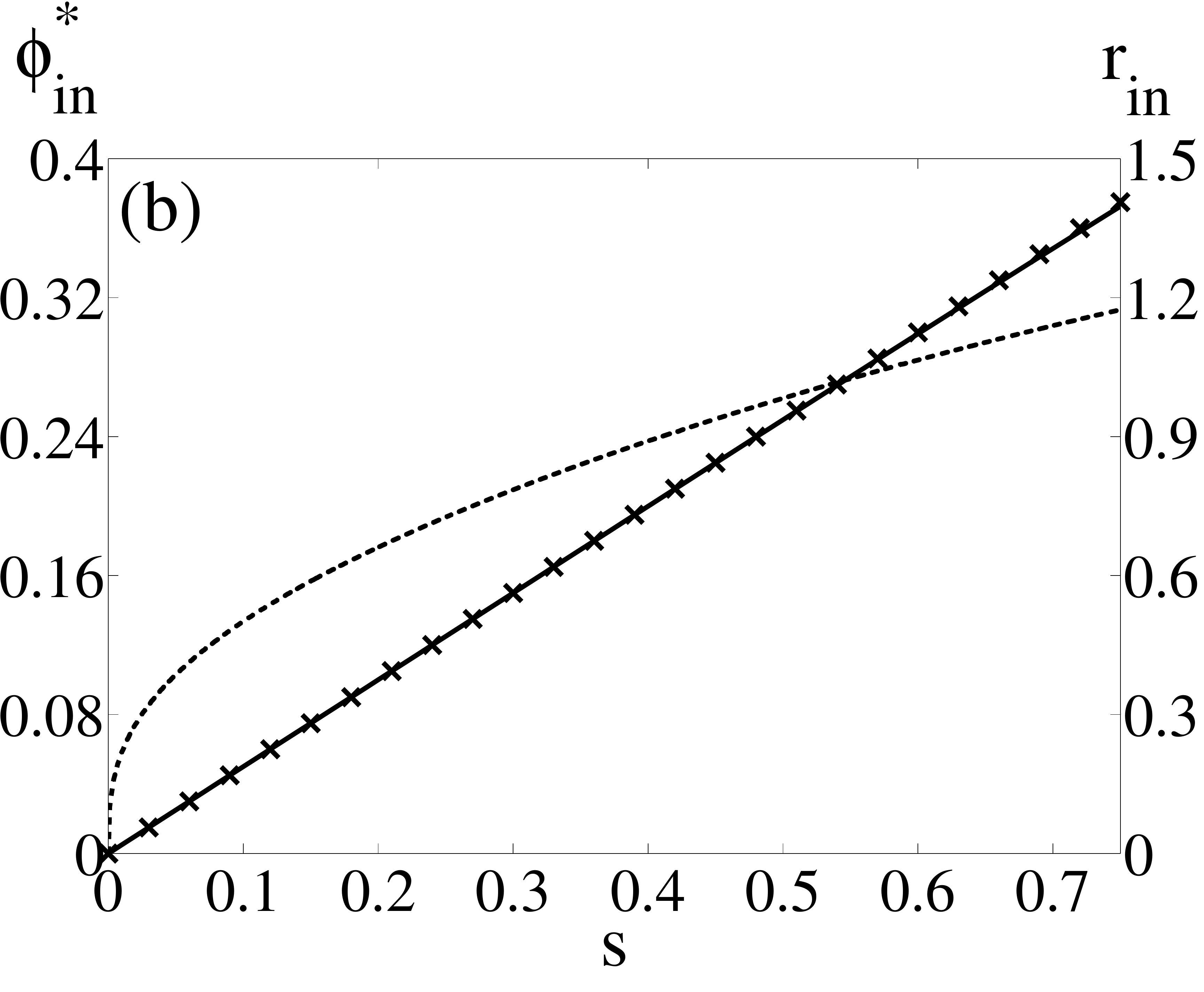}
	\end{minipage}
	\caption{Optimal input correlation $\phi_{\rm in}^*$ (solid line, left axis) and corresponding squeezing $r_{\rm in}$ (dashed line, right axis) vs. correlation ($\phi$ or $s$) for (a) the channel with additive Markov noise, where the crosses depict $\phi/2$; (b) the channel with non-Markovian noise (lossy and additive), where the crosses depict $s/2$. We took for both plots $N_M=N_N=1$ and $\overline{n}=5$.}
	\label{fig:optin}
\end{figure}
By comparing the spectrum of Eq. \eqref{eq:gmpsq} with the optimal input spectra [Eq. \eqref{eq:optin}] for the noise models of Eqs. \eqref{eq:envmark} and \eqref{eq:envoleg}, one can directly verify that the optimal input state is not a GMPS. However, one may use the GMPS as an approximation of the optimal input state for both these noise models. By calculating the transmission rates for noise models [Eqs. \eqref{eq:envmark} and \eqref{eq:envoleg}] with the GMPS as input state [using Eq.~\eqref{eq:capglwf} and replacing $\gamma_{\rm out}^{q,p*}(x)$ by $\gamma_{\rm GMPS}^{q,p}(x)+\gamma_{\rm env}^{q,p}(x)$], we find numerically that the highest transmission rate is achieved for a GMPS with \emph{nearest neighbor correlations} $\bm{\gamma}_{\rm GMPS,n.n.}$ \cite{AE06}. We find that among all GMPS given by Eq.~\eqref{eq:gmpsq}, which can be generated with the setup defined in Fig.~\ref{fig:gmps}, only the GMPS with nearest neighbor correlations has a symmetric spectrum, that is
\begin{equation}\label{eq:gmpsnnspec}
	\gamma_{\rm GMPS,n.n.}^q(x) = \gamma_{\rm GMPS,n.n.}^p(\pi-x).
\end{equation}
Since the noise spectra defined in Sec. \ref{sec:noises} satisfy the same symmetry, it is intuitively clear that this type of GMPS is the most suitable state for these noise models. The optical setup for the three-mode building block that generates this nearest-neighbor GMPS is depicted in Fig.~\ref{fig:gmps}(b). More details on it are provided in Sec. III C.

From Eq.~\eqref{eq:gmpsnnspec} and the fact that the GMPS used as an input is a pure state, i.e., $\gamma_{\rm GMPS}^q(x) \, \gamma_{\rm GMPS}^p(x) = 1/4, \forall x,$ we find that $\tilde{N} = (1+\phi_{\rm in}^2)/(1-\phi_{\rm in}^2)$ and $\tilde{N} \Delta = -1/2$. Thus, the nearest neighbor correlated GMPS has quadrature spectra
\begin{equation}\label{eq:gmpsnn}
	\gamma_{\rm GMPS,n.n.}^{q,p}(x) = \frac{1 + \phi_{\rm in}^2}{1 + \phi_{\rm in}^2 \mp 2\phi_{\rm in} \cos(x)} - \frac{1}{2},
\end{equation}
with the upper (lower) sign for the $q$ ($p$) quadrature. Therefore, when looking for the optimal transmission rate, one has to optimize only over the parameter $\phi_{\rm in}$. In order to satisfy the global water filling solution for the GMPS, we replace ${\gamma_{\rm in}^{{q}{*}}}(0)$ 
by ${\gamma_{\rm GMPS,n.n.}^{q}}(0)$ in Eq.~\eqref{eq:nthr}, which leads to a modified input energy threshold depending on $\phi_{\rm in}$, that is,
\begin{equation}\label{eq:nthrgmps}
	\overline{n}_{\rm thr}^{\rm GMPS} = \overline{n}_{\rm thr} - [{\gamma_{\rm in}^{{q}{*}}}(0) - \gamma_{\rm GMPS,n.n.}^{q}(0)].
\end{equation}
As we require that the input energy $\overline{n} \geq \overline{n}_{\rm thr}^{\rm GMPS}$, Eq.~\eqref{eq:nthrgmps} imposes an upper bound on $\phi_{\rm in}$.

In Figs.~\ref{fig:markrate}-\ref{fig:nonmarklossyrate}, we plot the rates obtained for the GMPS with the spectrum given by Eq.~\eqref{eq:gmpsnn} calculated via a maximization over $\phi_{\rm in}$, which we denote as $R_{\rm GMPS}$.
In Fig.~\ref{fig:markrate}, we observe that for the channel with additive Markov noise \eqref{eq:envmark}, $R_{\rm GMPS}$ is close-to-capacity achieving; in the plotted region, $R_{\rm GMPS}/C > 0.999$. For the additive channel with non-Markovian noise \eqref{eq:envoleg}, we conclude 
from Fig.~\ref{fig:nonmarkrate} that the GMPS serves as a very good resource as well; in the plotted region, $R_{\rm GMPS}/C > 0.999$. We confirm the same behavior for the lossy channel with non-Markovian noise, as shown in Fig.~\ref{fig:nonmarklossyrate} for different beamsplitter transmittances $\eta$. 

The optimal input correlations $\phi_{\rm in}^*$ for both noise models are approximately given by $\phi/2$ and $s/2$, respectively, as can be seen in Fig.~\ref{fig:optin}(a) and Fig.~\ref{fig:optin}(b). This can be verified as follows. Since the quantum water filling solution holds for the GMPS with nearest neighbor correlations, its rate is given by Eq.~\eqref{eq:capglwf} replacing $\gamma_{\rm out}^{q,p*}(x)$ by $\gamma_{\rm GMPS,n.n.}^{q,p}(x)+\gamma_{\rm env}^{q,p}(x)$. In order to find the optimal $\phi_{\rm in}$ it is sufficient to minimize only the second term in Eq.~\eqref{eq:capglwf} as only this term depends on $\phi_{\rm in}$. This term is a definite integral of a function whose primitive is not expressed in terms of elementary functions and $\phi_{\rm in}$. However, if the integrand as a function of parameter $\phi_{\rm in}$ can be properly minimized for all values of the variable of integration $x$ the integral will also be minimized. In order to verify this possibility we take the first derivative of the integrand and set it to zero. This leads to the following relation:
\begin{equation}\label{eq:optphiin}
	\frac{\gamma_{\rm env}^q(x)}{\gamma_{\rm env}^p(x)} = \frac{(1 + {\phi_{\rm in}^*}^2 + 2\phi_{\rm in}^* \cos x)^2}{(1 + {\phi_{\rm in}^*}^2 - 2\phi_{\rm in}^* \cos x)^2}.
\end{equation}
As it happens in the general case, there is no unique parameter $\phi_{\rm in}^*$ which satisfies Eq.~\eqref{eq:optphiin} for all $x$. Nevertheless, it is possible to obtain an approximating equality by neglecting the quadratic and higher order terms in the noise spectra given by Eqs.~\eqref{eq:envmark} and \eqref{eq:envoleg} and in the right-hand side of Eq.~\eqref{eq:optphiin}, i.e.
\begin{equation}
	\frac{1+2\alpha\cos(x)}{1-2\alpha\cos(x)} \approx \frac{1+4\phi_{\rm in}^*\cos(x)}{1-4\phi_{\rm in}^*\cos(x)}
\end{equation}
where $\alpha = \phi$ for the Markovian noise and $\alpha=s$ for the non-Markovian noise, respectively. This is a valid approximation taking into account that $\phi_{\rm in}^*<1$ and can be satisfied by a unique parameter $\phi_{\rm in}^*$ for all $x$. Namely, we find the simple relations $\phi_{\rm in}^* \approx \phi/2$ and $\phi_{\rm in}^* \approx s/2$, 
as verified in Fig.~\ref{fig:optin}(a) and Fig.~\ref{fig:optin}(b), respectively.

\subsection{GMPS as exact optimal input state}
Although we have seen that the GMPS is not the optimal input state for the noise models introduced in Sec. \ref{sec:noises}, it is possible to do better. Indeed, for all noises given by
\begin{equation}\label{eq:gmpsenv}
	\bm{\gamma}_{\rm env} = \left( \bm{\mathcal{N}}_{\rm env} \oplus \bm{\mathcal{N}}_{\rm env} \right) \times(\bm{\mathcal{C}}^{-1} \oplus \bm{\mathcal{C}}), 
\end{equation}
where $\bm{\mathcal{N}}_{\rm env}$ is an $n \times n$ matrix that commutes with $\bm{\mathcal{C}}$ given in Eq. \eqref{eq:gmps}, the GMPS 
is the \emph{exact} optimal input state, that is
\begin{equation}
	\bm{\gamma}_{\rm in}^* \equiv \bm{\gamma}_{\rm GMPS}, \quad \overline{n} \geq \overline{n}_{\rm thr}^{\rm GMPS},
\end{equation}
where now trivially $\overline{n}_{\rm thr}^{\rm GMPS}=\overline{n}_{\rm thr}$. This is a direct result that can be deduced from the shape of the 
CM $\bm{\gamma}_{\rm GMPS}$ 
and the fact that the CM of the optimal input state (given by Eqs.~\eqref{eq:optincm} and \eqref{eq:optin}) is diagonalized in the same basis as the CM of the noise.

Furthermore, as already mentioned, GMPS are known to be ground states of particular quadratic Hamiltonians \cite{SCW06}. More precisely, $\bm{\gamma}_{\rm GMPS}$ is the CM of the ground state of the translationary invariant Hamiltonian, given in natural units by
\begin{equation}\label{eq:hamiltonian}
	\bm{\hat{H}} = \frac{1}{2}\left(\sum\limits_i \hat{p}^2_i + \sum\limits_{i,j}\hat{q}_i \, V_{ij} \hat{q}_j\right),
\end{equation}
where $\hat{q}_i$ and $\hat{p}_i$ are the position and momentum operators of an harmonic oscillator at site $i$ and the potential matrix is simply given by $\bm{V} = \bm{\mathcal{C}}^2$, where $\bm{\mathcal{C}}$ is defined in Eq.~\eqref{eq:gmps}. 

A realistic example for a noise of the shape of Eq. \eqref{eq:gmpsenv} is given by the CM of the (Gaussian) state of the system defined in Eq.~\eqref{eq:hamiltonian}, i.e., a chain of coupled harmonic oscillators at finite temperature $T$. We assume the system to be described by a canonical ensemble, thus the density matrix of the oscillators is given by the Gibbs-state 
\begin{equation}
	\rho_{G} = \frac{\exp{(-\beta \bm{\hat{H}})}}{\mathrm{Tr}[\exp{(-\beta \bm{\hat{H}})}]},
\end{equation} 
where $\beta = 1/T$. The CM $\bm{\gamma}_G$ of the Gaussian state $\rho_G$ is given by Eq.~\eqref{eq:gmpsenv} with $\bm{\mathcal{N}}_{\rm env}=\bm{I} + [2\exp{(\beta \bm{\mathcal{C}})} - \bm{I}]^{-1}$ (see Ref.~\cite{AEPW02} for details), where indeed $[\bm{\mathcal{N}}_{\rm env},\bm{\mathcal{C}}]=0$. Therefore, if we assume the noise of the channel to result from a chain of coupled harmonic oscillators at finite temperature $T$, that is,
$\bm{\gamma}_{\rm env} = \bm{\gamma}_{G}$, then the GMPS with CM $\bm{\gamma}_{\rm GMPS}$ is both the ground state of the system given by Eq.~\eqref{eq:hamiltonian} and the {\it exact} optimal input state for $\overline{n} \geq \overline{n}_{\rm thr}$.

\subsection{Experimental realization}\label{sec:experiment}
Let us finally discuss the required optical squeezing strength to realize the optimal input correlation $\phi_{\rm in}^*$ for both noise models. We first present the mathematical description of the three-mode building block that generates the
GMPS with nearest neighbor correlations. The CM of this building block is given by \cite{AE07}
\begin{equation}\label{eq:gamma3mode}
	\bm{\gamma}_B = \frac{1}{2}
	\begin{pmatrix}
		w & v & u & 0   & 0   & 0\\
		v & w & u & 0   & 0   & 0\\
		u & u & t & 0   & 0   & 0\\
		0   & 0   & 0   & w & v & -u\\
		0   & 0   & 0   & v & w & -u\\
		0   & 0   & 0   &-u &-u & t
	\end{pmatrix},
\end{equation}
with $w=(t+1)/2$, $v=(t-1)/2$ and $u=\sqrt{(t^2-1)/2}$, where $t \ge 1$. The optical scheme for the three mode building block is depicted in Fig.~\ref{fig:gmps}(b), where $S(r_{B})$ is a one-mode squeezer with parameter $r_{B}$ such that $t = \cosh{(2r_{B})}$ \cite{AE07}. The resulting CM of the $n$-mode pure GMPS is given by \cite{AE06}
\begin{equation}
	\bm{\gamma}_{\rm GMPS} = \bm{\Gamma}_t - \trans{\bm{\Gamma}}_{wt}(\bm{\Gamma}_{ww}+ \theta \bm{\Gamma}_{\mathrm{TMS}} \theta)^{-1} \bm{\Gamma}_{wt},
\end{equation}
with $\theta=I\oplus -I$, where $I$ is the $n \times n$ identity matrix \footnote{We remark that the application of $\Theta$ on $\bm{\Gamma}_{\mathrm{TMS}}$ corresponds to a partial transpose $\hat{p}_i \rightarrow -\hat{p}_i$, which however has no effect here as $\bm{\Gamma}_{\mathrm{TMS}}$ does not contain any $q-p$ correlations.},
\begin{equation}
	\begin{split}
		\bm{\Gamma}_t & = \frac{1}{2}\bigoplus_{i=1}^{n} \mathrm{diag}\{t,t\},\\
		\trans{\bm{\Gamma}}_{wt} & = \frac{1}{2}\bigoplus_{i=1}^n \begin{pmatrix} u & u & 0 & 0\\ 0 & 0 & -u &-u\end{pmatrix},\\
		\bm{\Gamma}_{ww} & = \frac{1}{2}\bigoplus_{i=1}^{2n}\begin{pmatrix} w & v\\v & w \end{pmatrix},\\
		\bm{\Gamma}_{\mathrm{TMS}} & = \frac{1}{2}\bm{\gamma}_{\mathrm{TMS}}(r_T) \oplus \bm{\gamma}_{\mathrm{TMS}}(-r_T),		
	\end{split}
\end{equation}
where $\bm{\gamma}_{\mathrm{TMS}}(r)$
\begin{equation*}
	=
	\begin{pmatrix}
		\mathrm{ch}(r) 		& 0 				& 0			 		& \cdots 			& \cdots 			& \cdots 	& 0				& \mathrm{sh}(r)\\
		0 					& \mathrm{ch}(r) 		& \mathrm{sh}(r)		& 0 				& 0					& \cdots	& \cdots		& 0\\
		0					& \mathrm{sh}(r) 		& \mathrm{ch}(r) 		& 0 				& 0		 			& \cdots	& \cdots		& \vdots\\
		\vdots				& 0					& 0 				& \mathrm{ch}(r) 		& \mathrm{sh}(r) 		& 0			& \cdots 		& \vdots\\ 
		\vdots				& 0					& 0					& \mathrm{sh}(r) 		& \mathrm{ch}(r) 		& 0			& \cdots		& \vdots\\
		\vdots				& \vdots			& \vdots   	 		& \ddots     		& \ddots			& \ddots	& \ddots		& \vdots\\
		0					& \vdots			&  		   	 		& \ddots     		& \ddots			& 			& \ddots		& 0\\
		\mathrm{sh}(r)   		& 0					& \cdots			& \cdots			& 0					& \cdots	& 0				& \mathrm{ch}(r)
	\end{pmatrix},
\end{equation*}
where $\mathrm{ch}(r)=\cosh(2r)$ and $\mathrm{sh}(r)=\sinh(2r)$, respectively.

We observe that the nearest neighbor correlated GMPS requires only one squeezing parameter $r_{B}$ to generate the three-mode building block of Fig.~\ref{fig:gmps}(b). Furthermore, we can use finitely entangled TMS vacuum states with squeezing $r_T$. For simplicity, we set $r_{B}=r_T \equiv r_{\rm in}$ \footnote{This restriction still allows us to generate all possible input correlations $\phi_{\rm in}$.} and plot in Fig.~\ref{fig:optin} the squeezing strength needed to generate the optimal input correlation $\phi_{\rm in}^*$ for different noise correlations. For the Markov noise, in the plotted region the required correlation does not exceed $\phi_{\rm in, max}^* \approx 0.3$, which can be realized by $r_{\rm in, max}\approx 1.08$ (about $9.4$ dB squeezing). For the non-Markovian noise, the required correlation does not exceed $\phi_{\rm in, max}^* \approx 0.4$, which corresponds to $r_{\rm in, max} \approx 1.18$ (about $10.2$ dB squeezing). This shows that the required squeezing values for the presented setup could be realized with accessible non-linear media for a realistic assumption of noise correlations (these maximal squeezing values have recently been realized experimentally, see, e.g., Ref. \cite{MAESVS11}).

\section{Conclusions}
We have demonstrated that a one-dimensional Gaussian matrix-product state, a multimode entangled state which can be prepared sequentially, can serve as a very good approximation to the optimal
input state for encoding information into Gaussian bosonic memory channels. The fact that the GMPS can be prepared sequentially is crucial because it makes the channel encoding feasible,
progressively in time along with the subsequent uses of the channel. For the analyzed channels and noise models, the GMPS achieves more than 99.9\% of the Gaussian capacity and may be
experimentally realizable as the required squeezing strengths are achievable within present technology. Furthermore, we have introduced a class of channel noises, originating from a chain
of coupled harmonic oscillators at finite temperature, for which the GMPS is the \emph{exact} optimal multimode input state. Given that GMPS are ground states of particular quadratic
Hamiltonians, our findings could serve as a starting point to find useful connections between quantum information theory and quantum statistical physics.

\begin{acknowledgements}
J.S. is grateful to Antonio Ac\'in, Jens Eisert, and Alessandro Ferraro for helpful discussions, and
acknowledges a financial support from the Belgian FRIA foundation. The authors acknowledge financial
support from the Belgian federal government via the IAP research network Photonics$@$be, from the
Brussels Capital Region under project CRYPTASC, and from the F.R.S.-FNRS under project HIPERCOM.
\end{acknowledgements}


\end{document}